\documentclass[twocolumn,english]{aastex6}
\usepackage[T1]{fontenc}
\usepackage[latin9]{inputenc}
\usepackage{graphicx}

\makeatletter

\providecommand{\tabularnewline}{\\}

\let\jnl@style=\rm
\def\ref@jnl#1{{\jnl@style#1}}

\def\aj{\ref@jnl{AJ}}                   % Astronomical Journal
\def\actaa{\ref@jnl{Acta Astron.}}      % Acta Astronomica
\def\araa{\ref@jnl{ARA\&A}}             % Annual Review of Astron and Astrophys
\def\apj{\ref@jnl{ApJ}}                 % Astrophysical Journal
\def\apjl{\ref@jnl{ApJ}}                % Astrophysical Journal, Letters
\def\apjs{\ref@jnl{ApJS}}               % Astrophysical Journal, Supplement
\def\ao{\ref@jnl{Appl.~Opt.}}           % Applied Optics
\def\apss{\ref@jnl{Ap\&SS}}             % Astrophysics and Space Science
\def\aap{\ref@jnl{A\&A}}                % Astronomy and Astrophysics
\def\aapr{\ref@jnl{A\&A~Rev.}}          % Astronomy and Astrophysics Reviews
\def\aaps{\ref@jnl{A\&AS}}              % Astronomy and Astrophysics, Supplement
\def\azh{\ref@jnl{AZh}}                 % Astronomicheskii Zhurnal
\def\baas{\ref@jnl{BAAS}}               % Bulletin of the AAS
\def\bac{\ref@jnl{Bull. astr. Inst. Czechosl.}}
\def\caa{\ref@jnl{Chinese Astron. Astrophys.}}
\def\cjaa{\ref@jnl{Chinese J. Astron. Astrophys.}}
\def\icarus{\ref@jnl{Icarus}}           % Icarus
\def\jcap{\ref@jnl{J. Cosmology Astropart. Phys.}}
\def\jrasc{\ref@jnl{JRASC}}             % Journal of the RAS of Canada
\def\memras{\ref@jnl{MmRAS}}            % Memoirs of the RAS
\def\mnras{\ref@jnl{MNRAS}}             % Monthly Notices of the RAS
\def\na{\ref@jnl{New A}}                % New Astronomy
\def\nar{\ref@jnl{New A Rev.}}          % New Astronomy Review
\def\pra{\ref@jnl{Phys.~Rev.~A}}        % Physical Review A: General Physics
\def\prb{\ref@jnl{Phys.~Rev.~B}}        % Physical Review B: Solid State
\def\prc{\ref@jnl{Phys.~Rev.~C}}        % Physical Review C
\def\prd{\ref@jnl{Phys.~Rev.~D}}        % Physical Review D
\def\pre{\ref@jnl{Phys.~Rev.~E}}        % Physical Review E
\def\prl{\ref@jnl{Phys.~Rev.~Lett.}}    % Physical Review Letters
\def\pasa{\ref@jnl{PASA}}               % Publications of the Astron. Soc. of Australia
\def\pasp{\ref@jnl{PASP}}               % Publications of the ASP
\def\pasj{\ref@jnl{PASJ}}               % Publications of the ASJ
\def\rmxaa{\ref@jnl{Rev. Mexicana Astron. Astrofis.}}%
\def\qjras{\ref@jnl{QJRAS}}             % Quarterly Journal of the RAS
\def\skytel{\ref@jnl{S\&T}}             % Sky and Telescope
\def\solphys{\ref@jnl{Sol.~Phys.}}      % Solar Physics
\def\sovast{\ref@jnl{Soviet~Ast.}}      % Soviet Astronomy
\def\ssr{\ref@jnl{Space~Sci.~Rev.}}     % Space Science Reviews
\def\zap{\ref@jnl{ZAp}}                 % Zeitschrift fuer Astrophysik
\def\nat{\ref@jnl{Nature}}              % Nature
\def\iaucirc{\ref@jnl{IAU~Circ.}}       % IAU Cirulars
\def\aplett{\ref@jnl{Astrophys.~Lett.}} % Astrophysics Letters
\def\apspr{\ref@jnl{Astrophys.~Space~Phys.~Res.}}
\def\bain{\ref@jnl{Bull.~Astron.~Inst.~Netherlands}} 
\def\fcp{\ref@jnl{Fund.~Cosmic~Phys.}}  % Fundamental Cosmic Physics
\def\gca{\ref@jnl{Geochim.~Cosmochim.~Acta}}   % Geochimica Cosmochimica Acta
\def\grl{\ref@jnl{Geophys.~Res.~Lett.}} % Geophysics Research Letters
\def\jcp{\ref@jnl{J.~Chem.~Phys.}}      % Journal of Chemical Physics
\def\jgr{\ref@jnl{J.~Geophys.~Res.}}    % Journal of Geophysics Research
\def\jqsrt{\ref@jnl{J.~Quant.~Spec.~Radiat.~Transf.}}
\def\memsai{\ref@jnl{Mem.~Soc.~Astron.~Italiana}}
\def\nphysa{\ref@jnl{Nucl.~Phys.~A}}   % Nuclear Physics A
\def\physrep{\ref@jnl{Phys.~Rep.}}   % Physics Reports
\def\physscr{\ref@jnl{Phys.~Scr}}   % Physica Scripta
\def\planss{\ref@jnl{Planet.~Space~Sci.}}   % Planetary Space Science
\def\procspie{\ref@jnl{Proc.~SPIE}}   % Proceedings of the SPIE

\makeatother

\usepackage{babel}
\begin{document}

\title{Supernova and prompt gravitational-wave precursors to LIGO gravitational-wave
sources and short-GRBs }

\author{Erez Michaely$^{*}$ \& Hagai B. Perets}

\email{$^{*}$erezmichaely@gmail.com}

\affil{Physics Department, Technion - Israel Institute of Technology, Haifa
3200004, Israel}
\begin{abstract}
Binary black-holes (BHs) and binary neutron-stars (NSs) mergers had
been recently detected through gravitational-wave (GW) emission, with
the latter followed by post-merger electromagnetic counterparts, appearing
seconds up to weeks after the merger. While post-merger electromagnetic
counterparts had been anticipated theoretically, very little electromagnetic
precursors to GW-sources had been proposed, and non observed yet.
Here we show that a fraction of ${\rm a}\,{\rm few\times}10^{-4}-10^{-1}$
of LIGO GW-sources and short-GRBs, could be preceded by supernovae-explosions
years up to decades before the merger. Each of the BH/NS-progenitors
in GW-sources are thought to form following a supernova, likely accompanied
by a natal velocity-kick to the newly born compact object. The evolution
and natal-kicks determine the orbits of surviving binaries, and hence
the delay-time between the birth of the compact-binary and its final
merger through GW-emission. We use data from binary evolution population-synthesis
models to show that the delay-time distribution has a non-negligible
tail of ultra-short delay-times between 1-100 yrs, thereby giving
rise to potentially observable supernovae precursors to GW-sources.
Moreover, future LISA/DECIGO GW space-detectors will enable the detection
of GW-inspirals in the pre-mergers stage weeks to decades before the
final merger. These ultra-short delay-time sources could therefore
produce a unique type of promptly appearing LISA/DECIGO-GW-sources
accompanied by \emph{coincident} supernovae. The archival (and/or
direct) detection of precursor (coincident) SNe with GW and/or short-GRBs
will provide unprecedented characterizations of the merging-binaries,
and their prior evolution through supernovae and natal kicks, otherwise
inaccessible through other means. 
\end{abstract}

\section{Introduction}

\label{sec:Introduction}

The recent detection of gravitational-wave (GW; \citealt{Abbott2016,Abbott2016a,Abo+17})
mergers of double-compact-object binaries (DCO; including binary black-holes,
BBHs, and binary neutron-stars, BNSs) had opened a new era of GW-astronomy.
The joint detection of GWs and post-merger electromagnetic counterparts
(as in GW170817/GRB170817A) allows for unprecedented characterization
and exploration of compact objects, their GW inspiral and their potentially
explosive final merger. 

Extensive studies of the possible progenitors of such events gave
rise to a wide range of models and possible expectations for the rates
and properties of GW-mergers and their outcomes. The recent detection
of a short-GRB counterpart to the NS-NS merger GW170817/GRB170817A
provided the first direct confirmation for the origin of short-GRBs
from BNS mergers \citep{Paczynski1986,Goodman1986,Eichler1989}. Various
post-merger electromagnetic counterparts had been anticipated theoretically
and/or recently observed to occur seconds up to weeks after the merger
\citep[and references therein]{Abo+17}. However, electromagnetic
\emph{precursors} to GW-sources, not yet observed, had been little
explored, and are focus of the current study. such precursors could
provide unique information on the progenitor systems otherwise inaccessible
through other proposed/observed electromagnetic-counterparts.

Models for BH-BH, BH-NS and NS-NS mergers (detectable by current LIGO/VIRGO
and next generation GW detectors) explored the expected delay time
between the formation of the compact objects through supernovae (and/or
direct collapse) and the final detectable merger (e.g. \citealt{Belczynski2006a,Belczynski2008,Dominik2012}).
The delay\textendash time distribution (DTD) can provide valuable
information and constraints regarding the expected environment (e.g
star-forming vs. old) of the GW sources. Therefore DTD studies focused
on understanding the overall distribution and correspondence to Gyrs
old populations compared with young, up to tens of Myrs old populations.
Here we focus on the tail distribution of the \emph{shortest} delay
times and show that they give rise a small, but non-negligible fraction
of cases with ultra-short delay times (ranging from a year up to a
few decades) between the formation of the second compact object of
the binary progenitor and the final binary merger. Such ultra-short
delay sources have important implications for the existence of unique
type of supernova (SNe) precursor electromagnetic counterparts to
GW sources. The SNe arise from the final formation of the second compact-object
accompanying the GW source and preceding its final-merger by a year
to decades timescales. Note that though DTDs are extensively discussed
in the literature \citep[e.g.][]{Belczynski2006,Belczynski2007,Belczynski2008,Dominik2012,deMink2015},
their potential role as novel type of GW sources and preceding electromagnetic
counterparts, has only recently suggested by us \citep[hereafter paper I]{Mic+16}
and is explored here in details for the first time, to the best of
our knowledge. 

In addition, sufficiently short merger-times can result in a unique
novel type of GW sources, promptly appearing in the middle of the
detectable frequency band of next generation LISA/DECIGO GW detectors
\citep{Danzmann2003,Kawamura2006}, and accompanied by a concurrent
SN counterparts, as we discuss in the following.

The outline of this letter is as follows. We begin (section \ref{sec:Population-synthesis-models})
by describing the data and models for GW binary-progenitor we obtained
from publicly available population synthesis studies. We then (section
\ref{sec:Evolution-of-GW}) calculate the detailed GW-inspiral evolution
for each of the binary progenitors with ultra-short delay-times from
the available models, and obtain their GW strain and frequencies by
following their orbital properties as they evolve to the final merger.
We present our main results regarding the expected fractions and properties
of ultra-short delay-time events in section \ref{sec:Results} and
then discuss and summarize them in section \ref{sec:Discussion}.

\section{Population synthesis models and data}

\label{sec:Population-synthesis-models} 

We make use of data from the population synthesis study by \citet{Dominik2012},
openly available in the Synthetic Universe on-line database (www.syntheticuniverse.org)
from which we extracted the orbits of DCOs upon their initial formation.
\citet{Dominik2012} studied a wide range of models for the formation
and evolution of DCOs in the field and their merger rates (i.e. no
dynamical/tidal capture processes are taken into account). They used
the \texttt{StarTrack} population synthesis code \citep{Belczynski2002,Belczynski2008}
and explored the differences in the expected properties and their
dependence on the model assumptions and parameters. They focused on
some of the main uncertainties involved in binary stellar evolution,
and considered several models/parameters for the common envelope (CE)
phase; the maximal masses expected for NSs; the wind mass loss rate
prior the SN; and the natal kick for the NS/BH. They created two subsets
of 16 models named $A$ and $B$, which differ in their treatment
of the core-envelope transition problem (see details in \citealp{Belczynski2007}).
For each model they also considered two possible metalicities, $Z=Z_{\odot}$
and $Z=0.1Z_{\odot}.$ Altogether they explored $2\times2\times16=64$
models. For each model they followed the evolution of $10^{6}$ binaries
and extracted all the compact-binaries formed through the evolution,
including (1) BH-BH (2) BH-NS and (3) NS-NS binaries. 

The difference between subset models $A$ and $B$ due to the core-envelope
transition is an important one. It is not clear when in the evolution
of a star (late Hertzsprung gap (HG) or post-HG) does a clear boundary
appears between the low entropy core and the high entropy envelope.
This boundary is crucial for the outcome of the CE phase, if the companion
dissipated its orbital energy just on the envelope and survives as
close binary, or the entire star and merge. Submodel $A$ ignores
the core-envelope problem and just takes the energy balance into account,
while submodel $B$ assumes CE phase with a HG donors leads to a merger
and hence reduces the fraction of post-CE surviving binaries. 

The parameter $\lambda$ describes the binding energy of the envelope,
and is defined as:
\begin{equation}
E_{{\rm bind}}=-\frac{GM_{{\rm donor}}M_{{\rm donor,env}}}{\lambda R}
\end{equation}
where $R$ is the radius of the donor star. In \citet{Dominik2012}
the authors used seven different values of $\lambda$: four with fixed
values (V1-4), one is calculated as described in their section 2.3.2,
termed $\lambda_{{\rm Nanjing}}$ following \citet{Xu2010} and \citet{Loveridge2011}
(in their ``standard'' model) while in models V14 and V15 they vary
$\lambda_{{\rm Nanjing}}$ by a factor of 5 and 1/5 respectfully.
The merger rate also depends on the final NS mass; in V5 a maximal
mass of $M_{{\rm NS,max}}=3M_{\odot}$ is considered, while in V6
it is assumed that $M_{{\rm NS,max}}=2M_{\odot}$. Next the natal
kick issue is addressed in model V7; where the authors consider low-velocity
natal kicks, drawn from a Maxwellian distribution with velocity dispersion
of $\sigma=132.5{\rm kms^{-1}}$ for both NS and BH. In V8 they consider
high-velocity natal kick for the BH, while in model V9 the BHs are
not kicked at all. In V10 they use Delayed supernova engine (as compared
with the Rapid engine used in the standard model). Different winds
schemes are explored in model V11-V13. In V11 the mass-loss rates
are reduced by 50\% (compared with their standard model); in V12 they
assume fully conservative mass transfer; while V13 is fully non-conservative
mass transfer. A brief summary of all models considered can be in
Table \ref{tab:A-summary-table}; detailed explanations of each model
can be found in the original paper by \citet{Dominik2012}. 

\begin{table}
\begin{tabular}{|c|l|l|}
\hline 
Models & Parameter & Description\tabularnewline
\hline 
{\footnotesize{}S} & {\footnotesize{}Standard} & {\footnotesize{}$\lambda=$Nanjing, $M_{{\rm NS,max}}=2.5M_{\odot},$}\tabularnewline
 &  & {\footnotesize{}$\sigma=265{\rm kms^{-1}}$; BH kicks: variable }\tabularnewline
 &  & {\footnotesize{}SN: Rapid 0.5-cons. mass-transfer }\tabularnewline
\hline 
{\footnotesize{}V1} & {\footnotesize{}$\lambda=0.01$} & {\footnotesize{}very low $\lambda:$ fixed}\tabularnewline
\hline 
{\footnotesize{}V2} & {\footnotesize{}$\lambda=0.1$} & {\footnotesize{}low $\lambda:$ fixed}\tabularnewline
\hline 
{\footnotesize{}V3} & {\footnotesize{}$\lambda=1$} & {\footnotesize{}high$\lambda:$ fixed}\tabularnewline
\hline 
{\footnotesize{}V4} & {\footnotesize{}$\lambda=10$} & {\footnotesize{}very high$\lambda:$ fixed}\tabularnewline
\hline 
{\footnotesize{}V5} & {\footnotesize{}$M_{{\rm NS,max}}=3M_{\odot}$} & {\footnotesize{}high maximum NS mass}\tabularnewline
\hline 
{\footnotesize{}V6} & {\footnotesize{}$M_{{\rm NS,max}}=2M_{\odot}$} & {\footnotesize{}low maximum NS mass}\tabularnewline
\hline 
{\footnotesize{}V7} & {\footnotesize{}$\sigma=132.5{\rm kms^{-1}}$} & {\footnotesize{}low kicks: NS/BH}\tabularnewline
\hline 
{\footnotesize{}V8} & {\footnotesize{}full BH kick} & {\footnotesize{}high natal kicks: BH}\tabularnewline
\hline 
{\footnotesize{}V9} & {\footnotesize{}no BH kick} & {\footnotesize{}no natal kicks: BH}\tabularnewline
\hline 
{\footnotesize{}V10} & {\footnotesize{}Delayed SN} & {\footnotesize{}NS/BH formation:Delayed SN engine}\tabularnewline
\hline 
{\footnotesize{}V11} & {\footnotesize{}weak winds} & {\footnotesize{}Wind mass loss rates reduced to 50\%}\tabularnewline
\hline 
{\footnotesize{}V12} & {\footnotesize{}const. MT} & {\footnotesize{}Fully conservative mass transfer}\tabularnewline
\hline 
{\footnotesize{}V13} & {\footnotesize{}variable MT} & {\footnotesize{}Fully non-conservative mass transfer}\tabularnewline
\hline 
{\footnotesize{}V14} & {\footnotesize{}$\lambda\times5$} & {\footnotesize{}Nanjing $\lambda$ increased by 5}\tabularnewline
\hline 
{\footnotesize{}V15} & {\footnotesize{}$\lambda\times\frac{1}{5}$} & {\footnotesize{}Nanjing $\lambda$ decreased by 5}\tabularnewline
\hline 
\end{tabular}\caption{\label{tab:A-summary-table}A summary table with different models
properties taken from \citet{Dominik2012} their table 1.}
\end{table}

\section{Evolution of GW inspirals}

\label{sec:Evolution-of-GW}

Massive binary systems can potentially lead to the formation of DCOs.
A fraction of these systems will merge within a Hubble time via GW
emission. For a DCO system to form it needs to survive the binary
stellar evolution steps, including the CE phases (for sufficiently
small separations), when the stars grow in size to form giants; and
the two SNe in which the compact objects are thought to form (not-withstanding
direct collapse which may produce a silent or fainter transient event
when a BH is formed). Such evolution is considered in the various
population synthesis models. here we focus on the properties of the
formed DCO binaries and their final inspiral through GW-emission,
searching for sufficiently rapid inspirals leading to mergers on the
scales of at most 100 yrs. For this purpose, we followed the GW inspiral
of each of the DCO binaries extracted from the population synthesis
models, and derived their properties and their evolution in the strain-frequency
domain of current and future GW-detectors. 

If a DCO binary forms following the complex earlier stellar evolution,
it then evolves only through GW-emission. The equations that govern
the dynamics in a GW emitting systems are given by \citet{Pet64}
\begin{equation}
\frac{da}{dt}=-\frac{64}{5}\frac{G^{3}m_{1}m_{2}\left(m_{1}+m_{2}\right)}{c^{5}a^{3}\left(1-e^{2}\right)^{7/2}}\left(1+\frac{73}{24}e^{2}+\frac{37}{96}e^{4}\right)
\end{equation}
\begin{equation}
\frac{de}{dt}=-e\frac{304}{15}\frac{G^{3}m_{1}m_{2}\left(m_{1}+m_{2}\right)}{c^{5}a^{4}\left(1-e^{2}\right)^{5/2}}\left(1+\frac{121}{304}e^{2}\right)
\end{equation}
where $c$ is the speed of light and $G$ is Newton's constant. The
averaged rate at which a system dissipates its orbital energy is given
by \citet{Peters1963}
\begin{equation}
\left\langle \frac{dE}{dt}\right\rangle =-\frac{32}{5}\frac{G^{4}m_{1}^{2}m_{2}^{2}\left(m_{1}+m_{2}\right)}{c^{5}a^{5}\left(1-e^{2}\right)^{7/2}}\left(1+\frac{73}{24}e^{2}+\frac{37}{96}e^{4}\right)
\end{equation}
while $\left\langle \cdot\right\rangle $ indicates orbital averaging.
For circular orbits the entire energy is radiated in the second $n=2$
harmonic of the binary orbital frequency $f_{{\rm bin}}=\left(2\pi\right)^{-1}\left(G\left(m_{1}+m_{2}\right)/a^{3}\right)^{1/2}.$
For eccentric orbits the energy radiated in the nth harmonic is \citet{Peters1963}
\begin{equation}
\dot{E}_{n}\equiv\left\langle \frac{dE_{n}}{dt}\right\rangle =-\frac{32}{5}\frac{G^{4}m_{1}^{2}m_{2}^{2}\left(m_{1}+m_{2}\right)}{c^{5}a^{5}\left(1-e^{2}\right)^{7/2}}g\left(n,e\right)
\end{equation}
while $g\left(n,e\right)$ is the enhancement factor given by 

\begin{equation}
g\left(n,e\right)=\frac{n^{4}}{32}\bigg\{\bigg[J_{n-2}\left(ne\right)-2eJ_{n-1}\left(ne\right)+\frac{2}{n}J_{n}\left(ne\right)
\end{equation}

\[
+2eJ_{n+1}\left(ne\right)-J_{n+2}\left(ne\right)\bigg]^{2}+
\]
\[
\left(1-e^{2}\right)\left[J_{n-2}\left(ne\right)-2J_{n}\left(ne\right)+J_{n+2}\left(ne\right)\right]^{2}+\frac{4}{3n^{2}}J_{n}^{2}\left(ne\right)\bigg\}
\]
where $J_{n}\left(x\right)$ is the nth Bessel function. From the
equations above one can calculate the characteristic strain of the
GW signal for a specific distance. For sky orbit average \citet{Barack2004}
present the characteristic strain for the nth harmonic:

\begin{equation}
h_{{\rm c,n}}=\frac{1}{\pi R}\sqrt{\frac{2\dot{E}_{n}}{n\times\dot{f}}},\label{eq:H_c}
\end{equation}
where $R$ is the luminosity distance. Given eq. (\ref{eq:H_c}) we
can calculate the signal for a given binary system and determine whether
the signal is detectable given the sensitivity curve of a specific
GW-detector. For example, the green line-dash curve in Fig. (\ref{fig:Characteristic-ampltude-})
shows the evolution of a BH-NS binary system from the moment of formation
(the left-most point; lowest frequency; highest separation), until
the merger (right-most point; highest frequency; closest separation).
Considering the assumed distance to the GW-source and the system parameter
(see figure caption), one can determine the detailed evolution of
the binary, and hence the evolution of the characteristic strain and
frequency at any given point. The three black solid lines correspond
to the sensitivity curves of LIGO, DECIGO and LISA \citep{Larson2000,Yagi2011,Abbott2017},
in the specific example the BH-NS binary system is detectable in LISA/DECIGO
immediately as it forms and during the inspiral in DECIGO, while the
final merger itself is detectable only in LIGO. Using these tools
we can now characterize the overall properties of all binaries in
the full data sample for each of the models we analyze, as discuss
in the following.

\section{Results}

\label{sec:Results}

\begin{figure*}
\includegraphics[width=17cm]{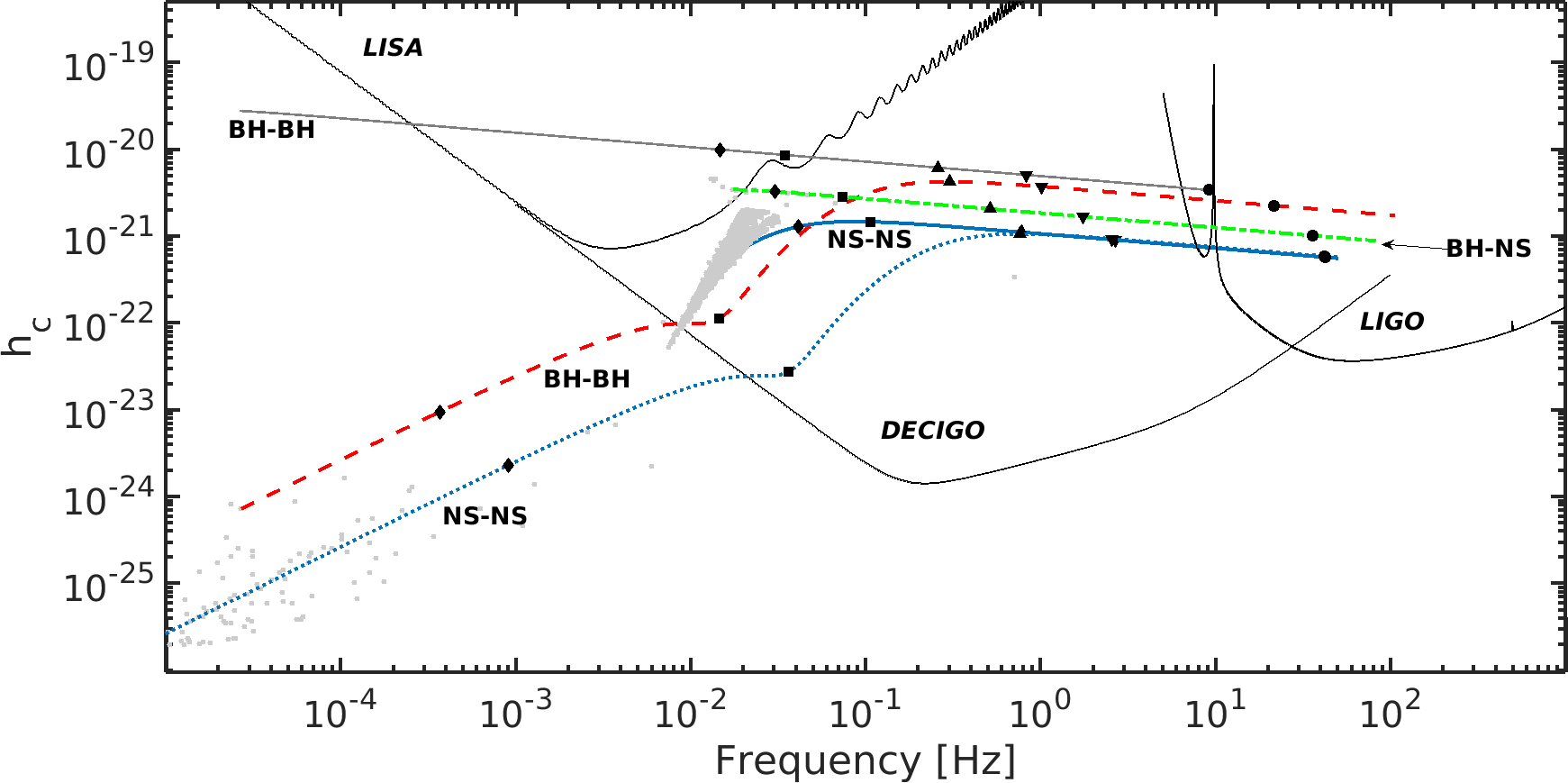}\caption{\label{fig:Characteristic-ampltude-}Characteristic evolution of ultra-short
delay-time binaries in the GW-strain ($h_{c})$-frequency domain,
compared with typical DCO binaries in the strain-frequency domain.
Five binary-evolution examples arising from submodel-A are shown,
with an assumed a luminosity distance of $R=50{\rm Mpc}$, with $Z=0.1Z_{\odot}$
metallicity. Grey thin-line depicts a reference typical BBH-system
which merges in $\sim200{\rm Myr}$. The other examples correspond
to ultra-short delay-time systems: (1) Blue solid-line: BNS evolution
(model V15;$_{\,}m_{1}=1.413M_{\odot};\ m_{2}=1.108M_{\odot};\ e=0.4772;\ a=0.047R_{\odot}$)
producing a prompt DECIGO-source with a coincident SN; (2) Blue dotted
line (V15;$\,m_{1}=1.3543M_{\odot};\ m_{2}=1.234M_{\odot};\ e=0.999;\ a=7.417R_{\odot}$).
(3) Red dashed-line: BBH evolution (V8;$\,m_{1}=5.736M_{\odot};\ m_{2}=5.605M_{\odot};\ e=0.9957;\ a=5.375R_{\odot}$;
(4) Green dashed-line: BH-NS evolution (V3;$\,m_{1}=5.626M_{\odot};\ m_{2}=1.159M_{\odot};\ e=0.0913;\ a=0.06R_{\odot})$
producing a prompt LISA-source. The three black solid-lines corresponds
to the sensitivity curves of LISA, DECIGO and LIGO. The black circles,
down-pointing triangles, up-pointing triangles, squares and diamonds
represent $1{\rm sec}$, $1{\rm hour}$, $1{\rm day}$, $1{\rm yr}$
and $10{\rm yr}$ before merger, respectfully. Gray-dots correspond
to the initial position of all the ultra-short systems that merge
within $100{\rm yr}$. All values are calculated assuming a luminosity
distance of $R=50{\rm Mpc}.$ }
\end{figure*}

\begin{figure*}
\includegraphics[width=17cm]{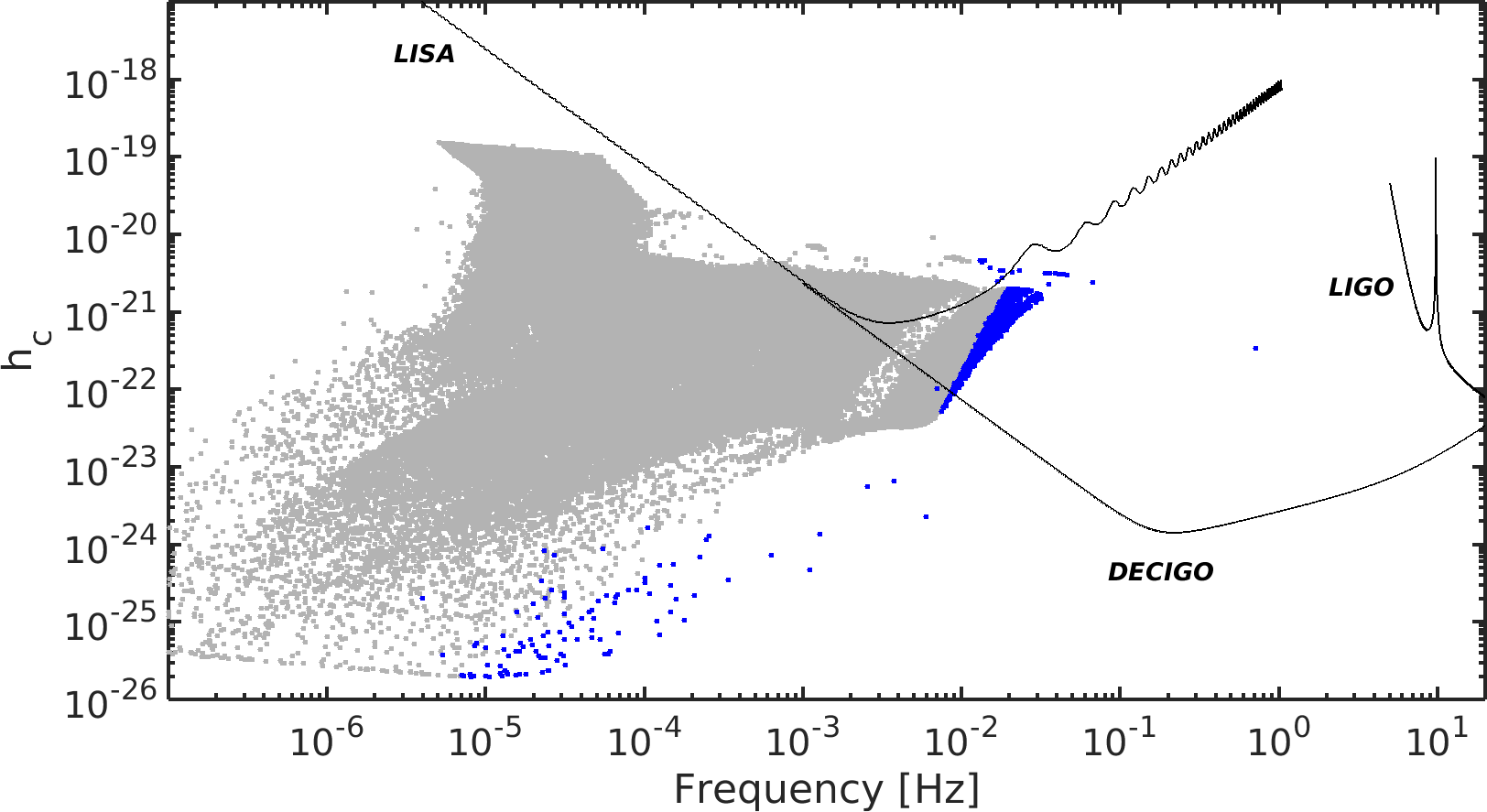}\caption{\label{fig:Prompt}The initial positions of all systems merging within
1 Gyr after the formation of the DCO binary. Blue dots indicate the
subgroup of ultra-short delay systems (see also Fig. 2) that merge
within $100{\rm yr}$. All data points are calculated with luminosity
distance $R=50{\rm Mpc}.$ We note that the characteristic strain
scales with $R^{-1}$ so for $R=500{\rm {\rm Mpc}}$ no prompt-GW
sources are expected in LISA but many are still observable as prompt-GW
sources in DECIGO.}
\end{figure*}

In order to better explain our detailed results, we first consider
a few specific cases that well-represent the various type of ultra-short
delay binaries we consider. In Fig. \ref{fig:Characteristic-ampltude-}
we plot five examples for the evolution of the characteristic strain
of several types of DCO systems. All examples are calculated for a
luminosity distance of $R=50{\rm Mpc}$ and originate from the type
A models with $Z=0.1Z_{\odot}$. A ``typical'' binary system is
depicted by a Grey thin line. This BH-BH binary begins outside the
LISA band, inspirals for $\sim200{\rm Myr}$ as it enters and crosses
the LISA and DECIGO bands, until it merges inside the LIGO band. The
other examples depict several examples from the sample of ultra-short
delay times binaries on which we focus in this work. Two NS-NS binaries
are shown. The first, depicted by the dotted Blue curve, forms following
a significant natal kick, and begins its evolution in an initially
highly-eccentric orbit (see detailed parameters in the figure caption).
This system is \textit{not} detectable immediately as it forms (i.e.
it is not a promptly appearing GW-source; see next section), but begins
its evolution outside the detection range. It then evolves, circularizes
and enter into the DECIGO sensitivity range on a short decades timescale
(a very similar evolution for a BH-BH binary is depicted by a red
dashed line). The second NS-NS binary, depicted by the solid Blue
curve, forms with with a moderate eccentricity, and promptly appears
inside the DECIGO detection band. Given its expected strain, would
be observable by DECIGO even up to $R\approx650{\rm Mpc.}$ The Green
line-dashed curve corresponds to a BH-NS binary promptly appearing
in the LISA detection band immediately as it forms, and then, just
a bit more than 10yrs later, it already merges inside the LIGO band,
likely producing a short-GRB. The Grey-dots show the initial position
of all the ultra-short delay-time binaries that merge within $100{\rm yrs}$
arising from all the models considered ($1688$ binaries in total).
Fig. \ref{fig:Prompt} shows all the ultra-short delay systems (within
100${\rm yr}$) marked in Blue; for comparison we also show \emph{all}
the other, much longer delay-time systems merging systems within $1{\rm Gyr}$
(Grey point; $368485$ systems in total). With this better understanding
of the type of ultra-short delay-time systems and their comparison
to other DCO systems, we are now ready to discuss the specific fractions
and distributions for the various models. 

\subsection{Delay time distributions of GW sources and short GRBs}

Let us denote the number of merging binaries at any given delay time
as $N_{{\rm i,j}}$ where subindex $i$ indicates the specific model
and subindex $j$ indicates the binary channel (BH-BH, BH-NS or NS-NS)
in the data set of \citet{Dominik2012}. For each merging binary the
full calculated evolutionary sequel provides us with the overall delay
time, as well relevant GW signal its frequency and strain (normalized
by the assumed distance to the source). 

\begin{figure*}
\includegraphics[width=1\columnwidth]{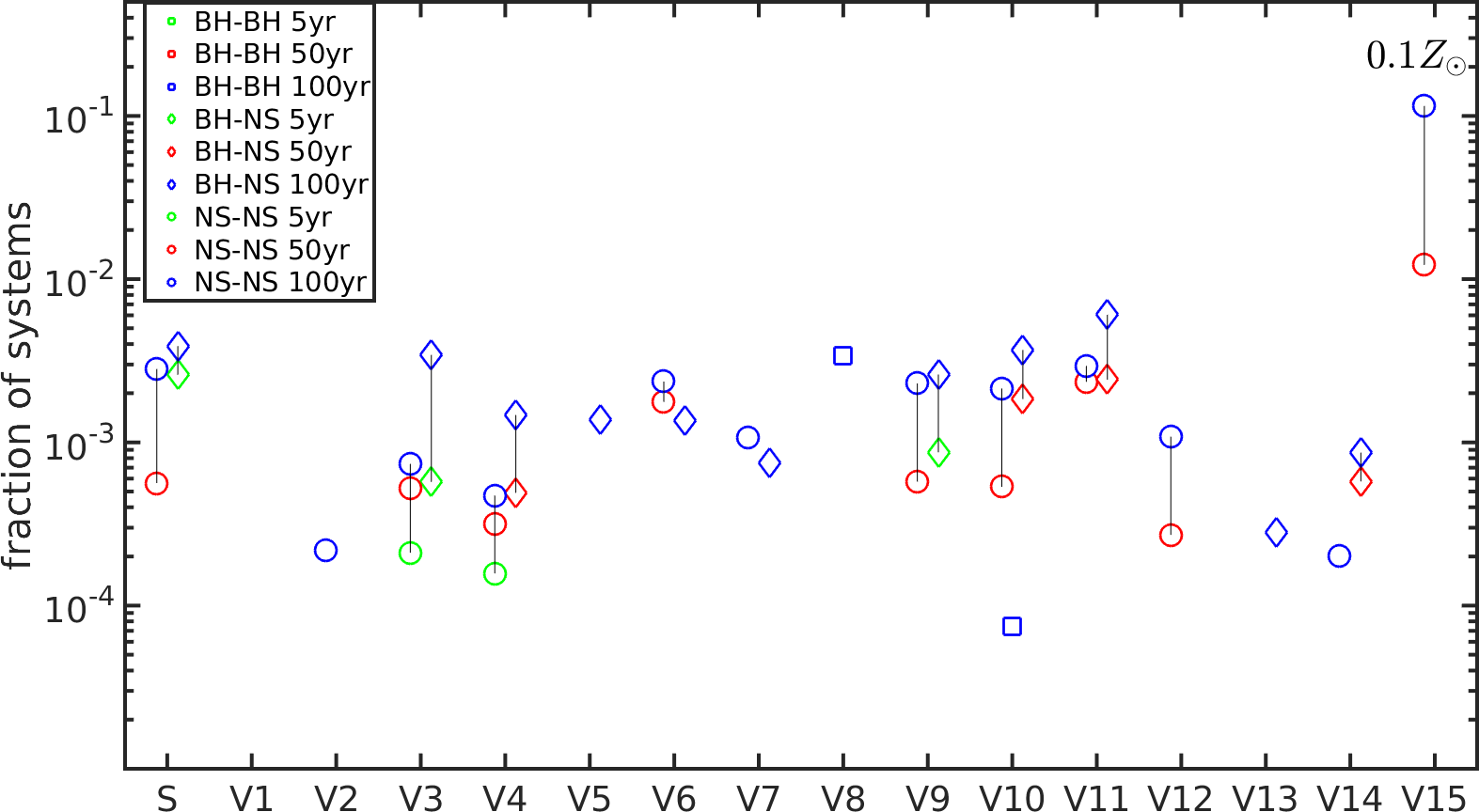}$\ $\includegraphics[width=1\columnwidth]{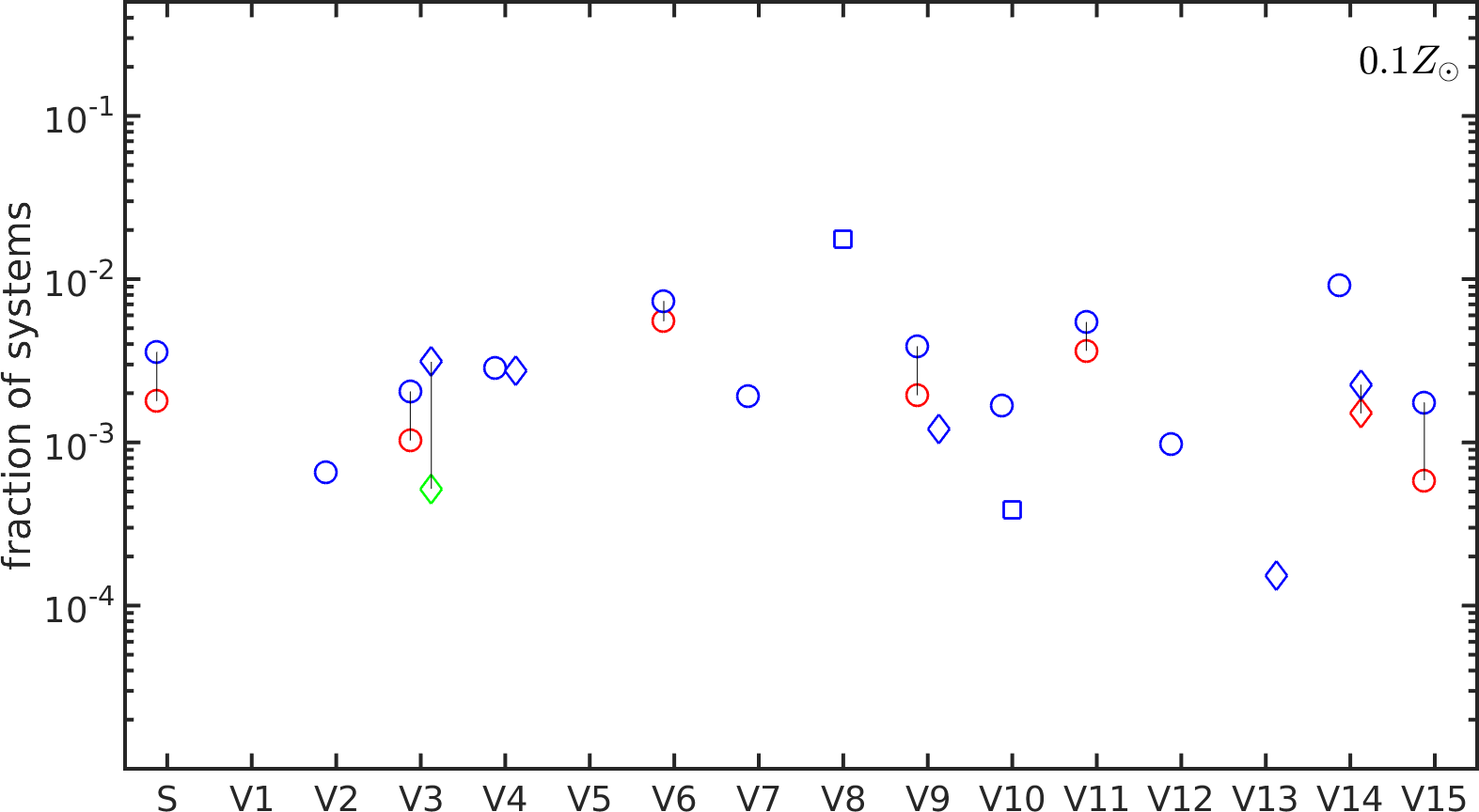}

\includegraphics[width=1\columnwidth]{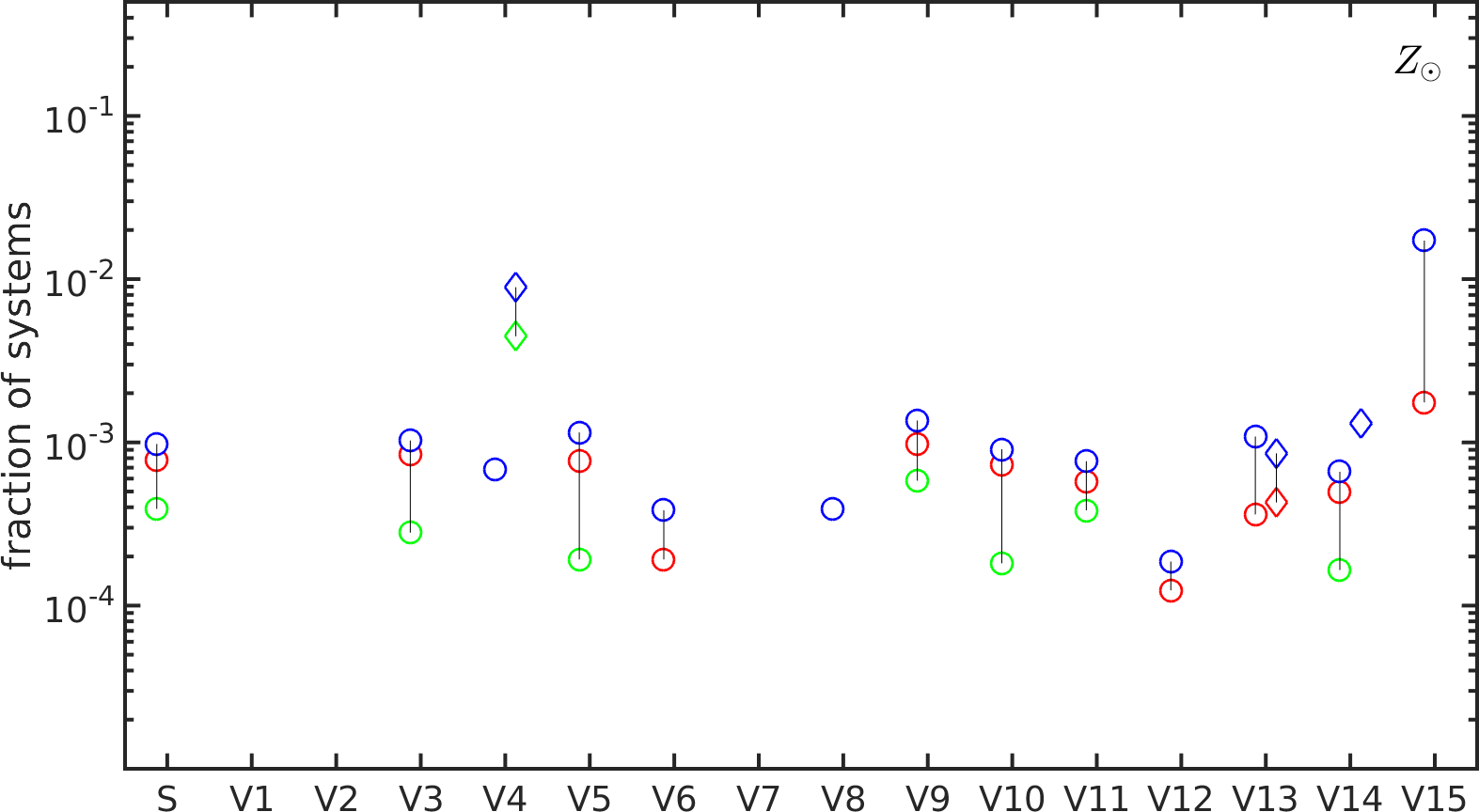}$\ $\includegraphics[width=1\columnwidth]{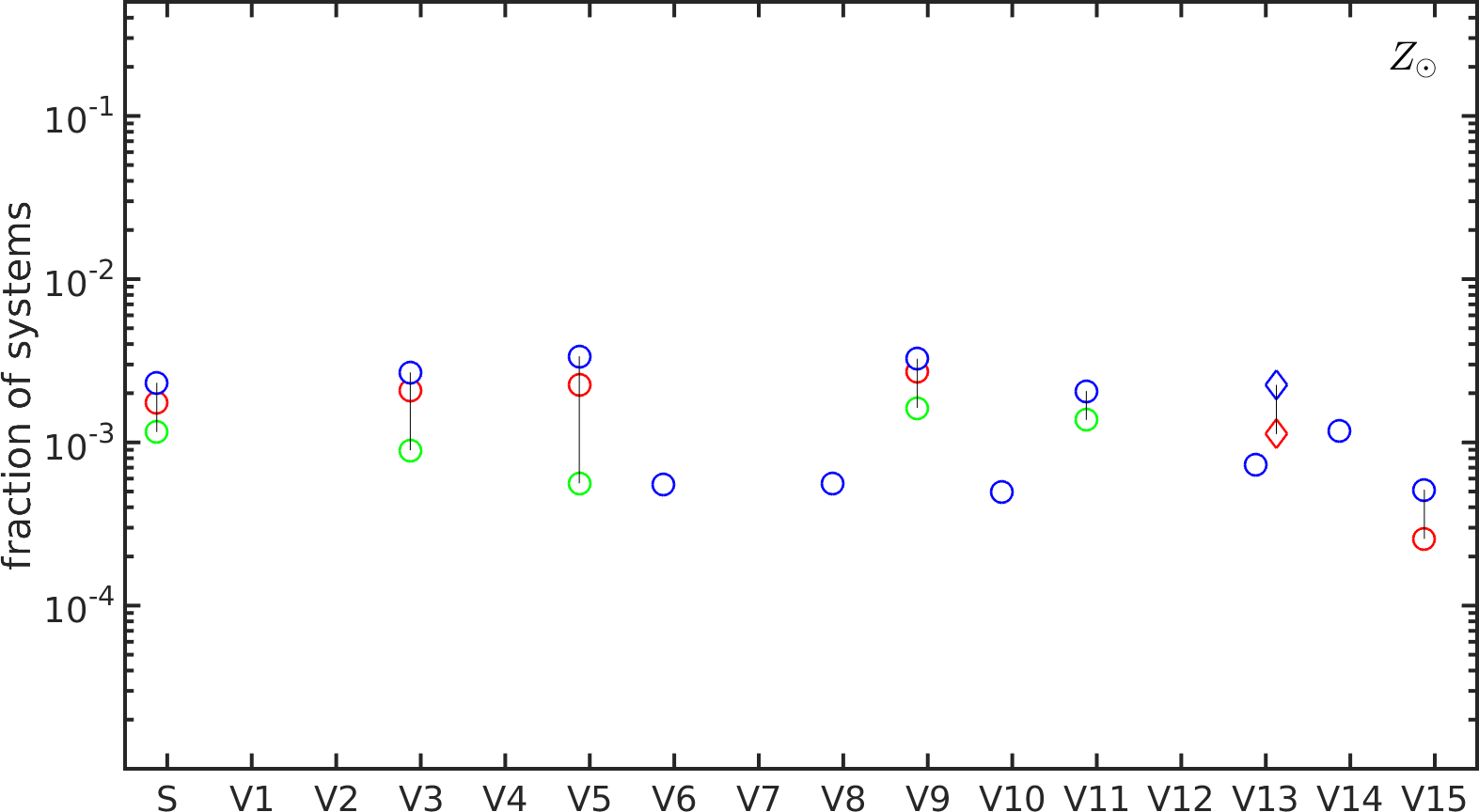}

\figcaption{\label{fig:Fraction-of-mergeringA}The fractions of merging systems
with ultra-short delay times. The markers correspond to the fractions
merging in 100, 10 and 5 yrs (from top to bottom; Blue, Red, Green,
respectively). Lines connect fractions arising from the same model
(horizontal axis). Left (right) plots correspond to Type A (B) models.
Upper (lower) plots correspond to $Z=Z_{\odot}$ ($Z=0.1Z_{\odot}$)
metallicity. The models parameters/assumptions are summarized in Table
1. }
\end{figure*}

Fig. \ref{fig:Fraction-of-mergeringA} shows the fraction of systems
that merge within some short-time $t_{{\rm thresh}}=5{\rm yr,}\ {\rm 50{\rm yr\ {\rm and}}\ 100yr}$
(marked by Green, Red and Blue markers, from bottom to top; respectfully,
linked by vertical lines) out of the total number of systems that
merge in less than 14 Gyrs ($N_{{\rm i,j}}$). BH-BH, BH-NS and NS-NS
binaries are represented by squares, diamonds and circles, respectfully.
Fig. \ref{fig:Fraction-of-mergeringA} corresponds to submodel A populations,
while Fig. \ref{fig:Fraction-of-mergeringA} corresponds to submodel
B populations. 

The majority of the models in Fig. \ref{fig:Fraction-of-mergeringA}
produce merging NS-NS/BH-NS binaries within $100{\rm yrs}$ with comparable
fractions ($5\times10^{-4}$ up-to $10^{-2}$) for solar-metallicity
models, and comparable, but lower fractions for the low-metallicity
models. Models V1 (solar-metallicity V2) and model V15 are the exception.
V1 does not produce any ultra-short delay systems due to the low-$\lambda$
parameter, leading to the merger of potential progenitors already
during the CE-phase. In contrast, model V15 (high-$\lambda$) produces
a large number of surviving-progenitors with $0.1$ of NS-NS systems
merging in $<100{\rm yrs}$. 

Interesting results also emerge from model V8, where no natal-kick
is given to the BH. We therefore expect to find a large fraction of
BH-BH binaries surviving the CE-phase and merging on short timescales.
The same analysis holds for submodel-B, (see right-column of Fig.
\ref{fig:Fraction-of-mergeringA}), showing only slightly lower-fractions
due to the smaller number of surviving binaries (due to the core-envelope
transition in this case). Note that in cases where no kicks are imparted,
the initial eccentricity of the binaries is lower, and they are more
likely to be detectable in \emph{both} LISA and DECIGO. Such systems,
however, have relatively large initial peri-center separation, and
without the kicks never attain initial sufficiently small separations
giving rise to the shortest delay-times. Indeed, though the fraction
of systems merging in 100yrs timescale is large, none of these systems
merge on $<10{\rm yrs}$. We refer the reader to Paper-I for discussion
and analytic treatment of the natal-kicks role in the formation of
ultra-short delay-time GW sources. 

\subsection{Prompt-GW-sources in LISA/DECIGO}

Future GW space-detectors such as the planned LISA and DECIGO missions
will be sensitive to the GW-signal arising from the early pre-merger
inspiral stage of DCO-binaries, weeks up to decades before the merger
\citep{Sesana2016,Chen2017}. Ultra-short delay-time mergers could
therefore appear promptly in middle of the LISA/DECIGO frequency-band,
coincident with a SN detection (e.g. Green dash-dotted and Blue-solid
lines in Fig. \ref{fig:Characteristic-ampltude-}). Such prompt appearance
differs from the typical long-delay sources that form at low-frequencies
outside the observable range and then evolve to higher frequencies
entering the LISA/DECIGO band, and very slowly moving to the right;
e.g. see the BH-BH Grey solid-line in Fig. \ref{fig:Characteristic-ampltude-}.
Such sources will therefore not be accompanied by a precursor/coincident
SN. The prompt-GW sources discussed here therefore constitute a novel
and unique type of sources with a unique signature. Note that at least
for close-by (50 Mpc) sources (see Fig. \ref{fig:Prompt}), many binaries
with longer delay times ($>100$ yrs) could also promptly-appear and
have a coincident SN, but will not reach the LIGO band during a human
life-time. 

\section{Discussion and summary}

\label{sec:Discussion}

In this \emph{letter }we studied the evolution and properties of double
compact object (NS-NS, NS-BH and BH-BH) binaries that inspiral and
merge on ultra-short timescales (years to decades following the initial
birth of the DCO-binary), typically following a SN explosion. The
short delay-times between the SN and the final merger, enable the
potential detection of the SN as an observable electromagnetic-precursor
to LIGO GW-sources and/or short-GRBs (the latter arising only from
NS-NS/NS-BH mergers). Since these binaries typically originate from
post-CE binaries with significantly-interacting binary progenitors,
the SN-precursor counterparts we identify are likely to be type Ib/c
SNe originating from massive stars that lost their outer envelopes
during the binary-evolution. Given the short delay-time, such events
will only be expected in star-forming regions/host-galaxies. Moreover,
given the short time elapsed since the second SN, the potentially
(natal-)kicked binary will not have sufficient time to further propagate
far from its birth-place in the galaxy, and such ultra-short delay
sources (and precursors) would be found at relatively small offsets
from their star-forming region birthplaces. 

The identification of a GW merger event in LIGO with a preceding SN
within year/decades would also provide a smoking-gun signature for
the binary-evolution origin of the DCO, rather than a dynamically-formed
GW-source, as the latter sources are not expected to have such precursors,
since they form long-after the formation of both DCO components. The
short delay-times also limit the possibility of ever identifying population
III ultra-short delay-time sources, as the SN-precursors of such systems
will explode at very high redshifts, and would likely be too faint
to be observable. 

Depending on the level of localization of the GW-sources, precursor
SNe could be paired with specific GW-sources at high probability given
a localization of the GW-source/short-GRB to a specific galaxy (given
the typical rate of $\sim10^{-2}$ yr$^{-1}$ for Milky-way-like galaxies),
but might become more challenging for poorly-localized GW-events,
given the background noise of SNe in a large ensemble of potential
host-galaxies. 

The fractions of expected sources with SN-precursors are sufficiently
high such that we might expect to find a few such events among the
hundreds to thousands of GW sources expected to be detected in the
coming few years, especially with the upgraded LIGO-VIRGO-KAGRA consortium.
Such identification would provide us with a direct link between a
SN that forms one of the secondary binary-component and the properties
of the compact object, and will potentially also inform us about the
type of binary that could form an ultra-short delay-time GW-source.
A statistical sample of such events would further constrain the evolutionary
models of DCO-binary progenitors. 

Finally, we proposed the existence of a novel type of promptly-appearing
GW-sources observable by next generation GW space-detectors. The induced
final stages of binary evolution and/or the natal kick given to the
second-formed compact object can produce an ultra-short delay-time
GW-source, which initial configuration would position it in the midst
of the LISA/DECIGO detection range, already upon their formation,
rather than the typical case of GW-sources entering the detection
range from lower frequencies. Moreover, such events could be coincidentally
observed with the SN. The coincidence would allow much better pairing
of the SN to the GW event, even with less than optimal localization
of the GW source, and such pairing would serve an important and unique
source of information about such events and their origins. 

\acknowledgements{We acknowledge support from the ISF-ICORE grant 1829/12.}

\bibliographystyle{apj}

\begin{thebibliography}{26}
\expandafter\ifx\csname natexlab\endcsname\relax\def\natexlab#1{#1}\fi

\bibitem[{{Abbott} {et~al.}(2016{\natexlab{a}}){Abbott}, {Abbott}, {Abbott},
  {Abernathy}, {Acernese}, {Ackley}, {Adams}, {Adams}, {Addesso}, {Adhikari},
  \& et~al.}]{Abbott2016a}
{Abbott}, B.~P., {Abbott}, R., {Abbott}, T.~D., {Abernathy}, M.~R., {Acernese},
  F., {Ackley}, K., {Adams}, C., {Adams}, T., {Addesso}, P., {Adhikari}, R.~X.,
  \& et~al. 2016{\natexlab{a}}, Physical Review Letters, 116, 241103

\bibitem[{{Abbott} {et~al.}(2016{\natexlab{b}}){Abbott}, {Abbott}, {Abbott},
  {Abernathy}, {Acernese}, {Ackley}, {Adams}, {Adams}, {Addesso}, {Adhikari},
  \& et~al.}]{Abbott2016}
---. 2016{\natexlab{b}}, Physical Review Letters, 116, 061102

\bibitem[{{Abbott} {et~al.}(2017{\natexlab{a}}){Abbott}, {Abbott}, {Abbott},
  {Abernathy}, {Ackley}, {Adams}, {Addesso}, {Adhikari}, {Adya}, {Affeldt}, \&
  et~al.}]{Abbott2017}
{Abbott}, B.~P., {Abbott}, R., {Abbott}, T.~D., {Abernathy}, M.~R., {Ackley},
  K., {Adams}, C., {Addesso}, P., {Adhikari}, R.~X., {Adya}, V.~B., {Affeldt},
  C., \& et~al. 2017{\natexlab{a}}, Classical and Quantum Gravity, 34, 044001

\bibitem[{{Abbott} {et~al.}(2017{\natexlab{b}}){Abbott}, {Abbott}, {Abbott},
  {Acernese}, {Ackley}, {Adams}, {Adams}, {Addesso}, {Adhikari}, {Adya}, \&
  et~al.}]{Abo+17}
{Abbott}, B.~P., {Abbott}, R., {Abbott}, T.~D., {Acernese}, F., {Ackley}, K.,
  {Adams}, C., {Adams}, T., {Addesso}, P., {Adhikari}, R.~X., {Adya}, V.~B., \&
  et~al. 2017{\natexlab{b}}, \apjl, 848, L13

\bibitem[{{Barack} \& {Cutler}(2004)}]{Barack2004}
{Barack}, L. \& {Cutler}, C. 2004, \prd, 69, 082005

\bibitem[{{Belczynski} {et~al.}(2002){Belczynski}, {Kalogera}, \&
  {Bulik}}]{Belczynski2002}
{Belczynski}, K., {Kalogera}, V., \& {Bulik}, T. 2002, \apj, 572, 407

\bibitem[{{Belczynski} {et~al.}(2008){Belczynski}, {Kalogera}, {Rasio}, {Taam},
  {Zezas}, {Bulik}, {Maccarone}, \& {Ivanova}}]{Belczynski2008}
{Belczynski}, K., {Kalogera}, V., {Rasio}, F.~A., {Taam}, R.~E., {Zezas}, A.,
  {Bulik}, T., {Maccarone}, T.~J., \& {Ivanova}, N. 2008, \apjs, 174, 223

\bibitem[{{Belczynski} {et~al.}(2006{\natexlab{a}}){Belczynski}, {Perna},
  {Bulik}, {Kalogera}, {Ivanova}, \& {Lamb}}]{Belczynski2006a}
{Belczynski}, K., {Perna}, R., {Bulik}, T., {Kalogera}, V., {Ivanova}, N., \&
  {Lamb}, D.~Q. 2006{\natexlab{a}}, \apj, 648, 1110

\bibitem[{{Belczynski} {et~al.}(2006{\natexlab{b}}){Belczynski}, {Sadowski},
  {Rasio}, \& {Bulik}}]{Belczynski2006}
{Belczynski}, K., {Sadowski}, A., {Rasio}, F.~A., \& {Bulik}, T.
  2006{\natexlab{b}}, \apj, 650, 303

\bibitem[{{Belczynski} {et~al.}(2007){Belczynski}, {Taam}, {Kalogera}, {Rasio},
  \& {Bulik}}]{Belczynski2007}
{Belczynski}, K., {Taam}, R.~E., {Kalogera}, V., {Rasio}, F.~A., \& {Bulik}, T.
  2007, \apj, 662, 504

\bibitem[{{Chen} \& {Amaro-Seoane}(2017)}]{Chen2017}
{Chen}, X. \& {Amaro-Seoane}, P. 2017, \apjl, 842, L2

\bibitem[{{Danzmann} \& {R{\"u}diger}(2003)}]{Danzmann2003}
{Danzmann}, K. \& {R{\"u}diger}, A. 2003, Classical and Quantum Gravity, 20, S1

\bibitem[{{de Mink} \& {Belczynski}(2015)}]{deMink2015}
{de Mink}, S.~E. \& {Belczynski}, K. 2015, \apj, 814, 58

\bibitem[{{Dominik} {et~al.}(2012){Dominik}, {Belczynski}, {Fryer}, {Holz},
  {Berti}, {Bulik}, {Mandel}, \& {O'Shaughnessy}}]{Dominik2012}
{Dominik}, M., {Belczynski}, K., {Fryer}, C., {Holz}, D.~E., {Berti}, E.,
  {Bulik}, T., {Mandel}, I., \& {O'Shaughnessy}, R. 2012, \apj, 759, 52

\bibitem[{{Eichler} {et~al.}(1989){Eichler}, {Livio}, {Piran}, \&
  {Schramm}}]{Eichler1989}
{Eichler}, D., {Livio}, M., {Piran}, T., \& {Schramm}, D.~N. 1989, \nat, 340,
  126

\bibitem[{{Goodman}(1986)}]{Goodman1986}
{Goodman}, J. 1986, \apjl, 308, L47

\bibitem[{{Kawamura} {et~al.}(2006)}]{Kawamura2006}
{Kawamura}, S. {et~al.} 2006, Classical and Quantum Gravity, 23, S125

\bibitem[{{Larson} {et~al.}(2000){Larson}, {Hiscock}, \&
  {Hellings}}]{Larson2000}
{Larson}, S.~L., {Hiscock}, W.~A., \& {Hellings}, R.~W. 2000, \prd, 62, 062001

\bibitem[{{Loveridge} {et~al.}(2011){Loveridge}, {van der Sluys}, \&
  {Kalogera}}]{Loveridge2011}
{Loveridge}, A.~J., {van der Sluys}, M.~V., \& {Kalogera}, V. 2011, \apj, 743,
  49

\bibitem[{{Michaely} {et~al.}(2016){Michaely}, {Ginzburg}, \&
  {Perets}}]{Mic+16}
{Michaely}, E., {Ginzburg}, D., \& {Perets}, H.~B. 2016, ArXiv:1610.00593

\bibitem[{{Paczynski}(1986)}]{Paczynski1986}
{Paczynski}, B. 1986, \apjl, 308, L43

\bibitem[{{Peters}(1964)}]{Pet64}
{Peters}, P.~C. 1964, Physical Review, 136, 1224

\bibitem[{{Peters} \& {Mathews}(1963)}]{Peters1963}
{Peters}, P.~C. \& {Mathews}, J. 1963, Physical Review, 131, 435

\bibitem[{{Sesana}(2016)}]{Sesana2016}
{Sesana}, A. 2016, Physical Review Letters, 116, 231102

\bibitem[{{Xu} \& {Li}(2010)}]{Xu2010}
{Xu}, X.-J. \& {Li}, X.-D. 2010, \apj, 716, 114

\bibitem[{{Yagi} \& {Seto}(2011)}]{Yagi2011}
{Yagi}, K. \& {Seto}, N. 2011, \prd, 83, 044011

\end{thebibliography}

\end{document}